 \documentclass[conference]{IEEEtran}
\IEEEoverridecommandlockouts
\usepackage{graphicx} 
\usepackage{epsfig} 
\usepackage{mathptmx} 
\usepackage{times} 
\usepackage{amsmath} 
\usepackage{amssymb}  
\usepackage{epstopdf}
\usepackage{subfigure}
\usepackage{setspace}
\usepackage{balance}
\usepackage{comment}
\usepackage{blkarray}
\usepackage{tikz}
\usetikzlibrary{intersections}

\numberwithin{subcase}{case}
\renewcommand{\arraystretch}{1.2}
\usetikzlibrary{arrows,shapes,positioning}
\usetikzlibrary{decorations.markings}
\tikzset{->-/.style={decoration={
  markings,
  mark=at position .5 with {\arrow{>}}},postaction={decorate}}}
\usepackage{booktabs}                     
\usepackage{multirow,rotating, array,booktabs}

\renewcommand\footnotemark{}

\title{Information Content in Neuronal Calcium Spike Trains: Entropy Rate Estimation based on Empirical Probabilities}

\makeatletter
\newcommand{\linebreakand}{%
  \end{@IEEEauthorhalign}
  \hfill\mbox{}\par
  \mbox{}\hfill\begin{@IEEEauthorhalign}
}
\makeatother

\author{\IEEEauthorblockN{Sathish Ande}
\IEEEauthorblockA{\textit{Department of Electrical Engineering} \\ \textit{IIT Hyderabad}, Hyderabad, India
}

\and
\IEEEauthorblockN{Srinivas Avasarala}
\IEEEauthorblockA{\textit{Department of Electrical Engineering} \\ \textit{IIT Hyderabad}, Hyderabad, India
}
\and
\IEEEauthorblockN{Jayanth R Regatti}
\IEEEauthorblockA{\textit{Department of Electrical Engineering} \\ \textit{IIT Hyderabad}, Hyderabad, India 
}
\and
\IEEEauthorblockN{Neha Pandey}
\IEEEauthorblockA{\textit{Department of Electrical Engineering} \\ \textit{IIT Hyderabad}, Hyderabad, India
} 
\and
\IEEEauthorblockN{Sarpras Swain}
\IEEEauthorblockA{\textit{Department of Chemical Engineering} \\ \textit{IIT Hyderabad}, Hyderabad, India 
}
\and
\IEEEauthorblockN{Ajith Karunarathne}
\IEEEauthorblockA{\textit{Department of Chemistry and Biochemistry}
\\\textit{The University of Toledo},
Ohio, USA\\
}
\linebreakand 
\IEEEauthorblockN{Lopamudra Giri}
\IEEEauthorblockA{\textit{Department of Chemical Engineering} \\ \textit{IIT Hyderabad}, Hyderabad, India 
}
\and
\IEEEauthorblockN{Soumya Jana}
\IEEEauthorblockA{\textit{Department of Electrical Engineering} \\ \textit{IIT Hyderabad}, Hyderabad, India 
}
}

\begin{document} 

\maketitle

\begin{abstract}
Quantification of information content and its temporal variation in intracellular calcium spike trains in neurons helps one understand functions such as memory, learning, and cognition. Such quantification could also reveal pathological signaling perturbation that potentially leads to devastating neurodegenerative conditions including Parkinson's, Alzheimer's, and Huntington's diseases. Accordingly, estimation of entropy rate, an information-theoretic measure of information content, assumes primary significance. However, such estimation in the present context is challenging because, while entropy rate is traditionally defined asymptotically for long blocks under the assumption of stationarity, neurons are known to encode information in short intervals and the associated spike trains often exhibit nonstationarity. Against this backdrop, we propose an entropy rate estimator based on empirical probabilities that operates within windows, short enough to ensure approximate stationarity. Specifically, our estimator, parameterized by the length of encoding contexts, attempts to model the underlying memory structures in neuronal spike trains. In an example Markov process, we compared the performance of the proposed method with that of versions of the Lempel-Ziv algorithm as well as with that of a certain stationary distribution method and found the former to exhibit higher accuracy levels and faster convergence. Also, in experimentally recorded calcium responses of four hippocampal neurons, the proposed method showed faster convergence. 
Significantly, our technique detected structural heterogeneity in the underlying process memory in the responses of the aforementioned neurons. We believe that the proposed method facilitates large-scale studies of such heterogeneity, which could in turn identify signatures of various diseases in terms of entropy rate estimates.

\end{abstract}


\section{INTRODUCTION}
\label{sec:intro}
Neurons are known to encode stimulus information in temporal spiking patterns of action potentials, which convey information about high-level functions including learning and memory \cite{reinagel2000temporal, collell2015brain}. In this context, measuring information content of neural spike trains and tracking its variation with time assume significance. However, estimation of entropy rate, the traditional measure of information content \cite{cover2012elements}, faces certain practical difficulty.
While entropy rate estimation for stationary processes is not difficult, 
neuronal spike trains often exhibit nonstationarity, and the challenge lies in accurately capturing temporal variation in the entropy rate \cite{vu2009information}.
To meet this challenge, we take inspiration from speech analysis \cite{tyagi2006variable}, assume window-wise stationarity, and estimate the local entropy rate. However, while local stationarity may be assumed within a small window, one requires a large window for the entropy rate estimate to converge to a reliable value \cite{strong1998entropy}.
So, the task boils down to choosing an entropy rate estimator that is accurate yet converges quickly, and pairing such an estimator with suitable window size, neither too long, nor too short. In this paper, we propose an entropy rate estimator based on empirical probabilities, and  experimentally demonstrate its superiority over competing methods on a theoretical example, as well as on calcium spike trains observed in neurons extracted from the hippocampus, a region of the brain associated with learning and memory \cite{collell2015brain}.    

\begin{figure}[t!]
\centering
\includegraphics[width=0.55\columnwidth,height=0.1\textheight]{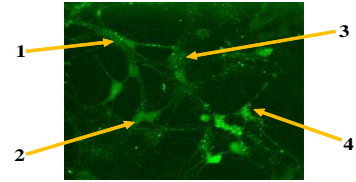}
\label{fig:calciumImaging}
\caption{Intracellular calcium imaging: Representative image of hippocampal neuron population with four neurons indexed 1-4, as explained in Section \ref{ssec:spikeTrain}. Scale bar = 20 $\mu$m.}
\label{fig:Calcium Imaging}
\end{figure}

Early research in this direction assumed independent and identically distributed ({\em iid}) spiking process, and unrealistically ignored timing information and possible memory \cite{jeong2001mutual,dilorenzo2013spike}.
Subsequently, as entropy rate estimator, 
versions of Lempel-Ziv (LZ) algorithm were suggested \cite{amigo2004estimating}, which are guaranteed to converge for large block length $n$ irrespective of memory structure \cite{effros2002universal}. Block-based methods were also proposed for large $n$ 
\cite{strong1998entropy}. Advanced methods have modeled the spiking process as stationary and Markov \cite{nakahama1983markov}. One such method used finite context trees up to certain depth $k$ and assigned weights to those trees  \cite{kennel2005estimating}; however, such weights, based on {\em ad hoc} assumptions, could be unreliable. Another method, assuming hierarchical Dirichlet priors, was demonstrated only for large block length $n$ \cite{knudson2013spike}. Interestingly, short blocks were recently considered under first and second order Markov models (albeit not in a neuronal study) \cite{vegetabile2019estimating}; however,
estimated higher order transition probabilities based on stationary distributions could be error prone for higher orders. In contrast, we propose an entropy rate estimation method that estimates empirical conditional distributions and generalizes to higher order Markov models. 

\usetikzlibrary{shapes.geometric, arrows}
\tikzstyle{startstop} = [rectangle, rounded corners, minimum width=3cm, minimum height=0.8cm,text centered, draw=black]
\tikzstyle{io} = [trapezium, trapezium left angle=70, trapezium right angle=110, minimum width=3cm, minimum height=1.5cm, text centered, draw=black, fill=blue!30]
\tikzstyle{process} = [rectangle, minimum width=3cm, minimum height=1cm, text centered, draw=black, fill=orange!30]
\tikzstyle{decision} = [diamond, minimum width=3cm, minimum height=1cm, text centered, draw=black, fill=green!30]
\tikzstyle{arrow} = [thick,->,>=stealth]

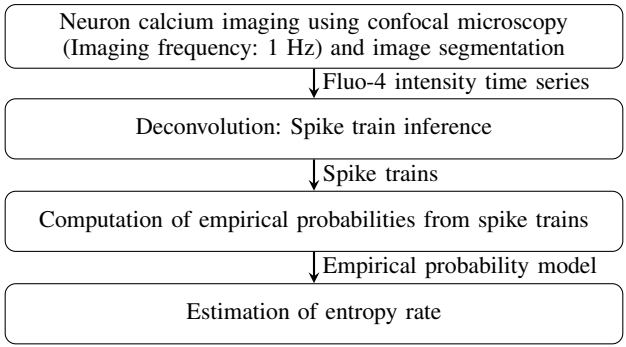
\begin{figure}[t!]
    \centering
   \small
 \begin{tikzpicture}[node distance=35pt]
\node (1) [startstop,text width = 8cm] {Neuron calcium imaging using confocal microscopy (Imaging frequency: 1 Hz) and image segmentation};
\node (2) [startstop,below of =1,text width = 8cm] {Deconvolution: Spike train inference};
\node (3) [startstop,below of =2,text width = 8cm] {Computation of empirical probabilities from spike trains};
\node (4) [startstop,below of =3,text width = 8cm] {Estimation of entropy rate};
\draw [arrow] (1) --node [anchor=west] { Fluo-4 intensity time series} (2) ;
\draw [arrow] (2) --node [anchor=west] {Spike trains}(3);
\draw [arrow] (3) -- node [anchor=west] {Empirical probability model}(4);
\end{tikzpicture}
    \caption{Schematic depiction of workflow.}
    \label{fig:workflow}
\end{figure} 

Historically, similar issue of statistical modeling also arose in text compression, where LZ algorithms were found ineffective in compressing short texts \cite{cover2012elements}. In that context, prediction by partial matching (PPM), based on empirical Markov modeling, proved effective \cite{cleary1997unbounded}. However, PPM, while closely related to entropy rate estimation, uses certain escape symbols to allow decompression, which is irrelevant for present purpose.
Accordingly, we propose an algorithm that makes use the main principle behind PPM, but excludes escape symbols.
The rest of this paper is organized as follows. Section~\ref{sec:materials} describes collection of calcium imaging data and the deconvolution technique for spike train inference, introduces the notion of entropy rate of stochastic process, and elaborates LZ algorithms and proposed empirical entropy rate estimator. Further, Section~\ref{sec:results} presents the results demonstrating superiority of the proposed method. Finally, Section~\ref{sec:conclusion} concludes the paper.

\section{MATERIALS AND METHODS}
\label{sec:materials}


The workflow of this paper is shown in Fig. \ref{fig:workflow}, and elaborated in the following.

\subsection{Data Collection}
\label{ssec:calcium Imaging}

Hippocampal neurons were cultured from l day postnatal Sprague-Dawley rats. We performed time-lapse imaging for monitoring intracellular calcium in hippocampal neurons at $7^{th}$ day after plating. The neurons were loaded with 1 $\mu$M Fluo-4 (Molecular Probes) for 30 min in Hank’s Balanced Salt Solution (HBSS).  The cells were then washed with HBSS for 3 times and Fluo-4 intensity was monitored using excitation at 488 nm and emission at 510 nm. We performed imaging using a spinning-disk confocal imaging system comprising a Leica DMI6000B inverted microscope, a Yokogawa CSU-X1 spinning-disk units \cite{giri2014g}. The interval between successive images, while set at 1 s, was observed to be between 0.8 s to 1 s due to inherent variabilities. During imaging, neurons were kept in the incubation chamber with the microscope maintained at 37$^o$C and 5\% CO$_2$. 
\begin{figure}[t!]
\centering
\vspace{-1.1em}
\includegraphics[width=\columnwidth,height=0.2\textheight]{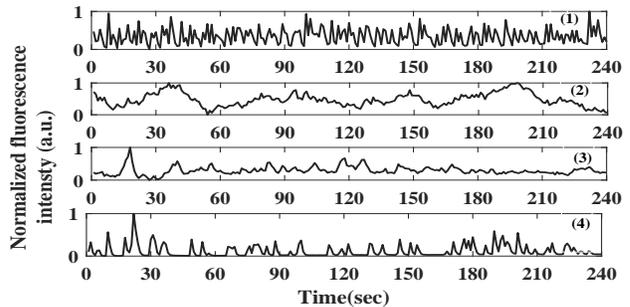}
 \vspace{-3.4em}
\caption{Time course of normalized Fluo-4 intensity for four neurons indexed 1-4 in Fig: \ref{fig:calciumImaging}, as explained in Section \ref{ssec:spikeTrain}.}
\label{fig:fluorescence-2}
\end{figure}

\begin{table*}[t!]
\centering
\renewcommand*{\arraystretch}{1.1} 
\caption{Empirical probability estimation}
\resizebox{0.8\textwidth}{!}{
{\small
  \begin{tabular}{ccccccccccccc}
  \toprule
 \multicolumn{3}{c}{$k=0$ (no context)}  && \multicolumn{4}{c}{$k=1$} && \multicolumn{4}{c}{$k=2$} \\
  \midrule
\multirow{ 2}{*}{Symbol}   & \multirow{ 2}{*}{Count}    &  Relative   && \multirow{ 2}{*}{Context}    & \multirow{ 2}{*}{Symbol}   & \multirow{ 2}{*}{Count}    &  Relative    &&  \multirow{ 2}{*}{Context}    & \multirow{ 2}{*}{Symbol}    &  \multirow{ 2}{*}{Count}   &  Relative  \\
&    &   frequency    &&    &    &  &  frequency   &&  &  & &  frequency     \\ \cline{1-13}
 0& 11   & 11/20  &&    0&0    &4  &2/5    &&   00& 0  & 1   & 1/3    \\ 
 1 & 9   & 9/20 &&  0& 1 &6   & 3/5    &&    00& 1  & 2   & 2/3    \\ 
 & &     && 1 &0  & 6    & 2/3   &&   01& 0  & 3   & 1/2     \\
   &   &  && 1   &1    &3  & 1/3    && 01 & 1  & 3   & 1/2  \\
  &   &    &&    &    &   &    && 10& 0  & 3   & 1/2 \\
      &    &  &&  &  &    &    && 10& 1  & 3   & 1/2 \\
  &    &   &&  &  &    &    && 11& 0  & 3   & 1   \\
  &    &   &&    &    &   &    && 11& 1  & 0   & 0    \\
  \bottomrule
  \end{tabular}}}
 \label{table:Empirical_Probability_model}%
\end{table*}%

\subsection{Spike Train Inference}
\label{ssec:spikeTrain}
The time course of spatially resolved Fluo-4 fluorescence intensity in neuron populations were obtained using Andor software from the time-lapse image data (see Fig. \ref{fig:Calcium Imaging} for representative time-lapse image). Regions of interest were marked at subcellular levels (soma regions). Further, intensity was calculated by subtracting the background intensity level, which was then normalized \cite{swain2017spatially}.
See Fig. 
\ref{fig:fluorescence-2} for representative time course of normalized intensity of four neurons indexed in Fig.\ref{fig:Calcium Imaging}. To infer spike train from time course of normalized fluorescence intensity, we used a fast nonnegative deconvolution algorithm  \cite{vogelstein2010fast}. 
In this algorithm, the estimated spike rate should be interpreted as the expected number of spikes in the 1s imaging interval around each time point. Unfortunately, such estimated spiking rates are generally noisy, and may lead to erroneous inference. Such errors were mitigated by setting an adaptive threshold
\begin{equation}  
P_{th} = 3 \sigma_p, 
\label{eq:threshold}
\end{equation}
$\sigma_p$ indicating the standard deviation of inferred spiking rate vector \cite{patel2015automated}. Specifically, the instants were labeled 1, when the spiking rate exceeded $P_{th}$ (high spiking), and 0, otherwise (low spiking). Such binary spike sequences were used for further analysis.

\subsection{Mathematical Preliminaries} 
\label{sec:math}

At this point, we introduce the notion of entropy rate, and describe existing entropy rate estimators. 

\subsubsection{Entropy rate}
\label{ssec:Entropyrate}

Entropy $H(Y)$, a measure of uncertainty, of random variable $Y$ with probability mass function $p_Y$ defined on alphabet ${\mathcal{Y}}$ is defined by $H(Y)=-\sum_{y\in \mathcal{Y}} p_Y(y)\log p_Y(y)$. Further, the entropy rate  $\bar{H}(X)$ of a sequence of random variables $X_1,X_2,\hdots,X_n,\hdots$, is defined by
\begin{align} \label{eq: entropyRate}
\bar{H}(X)=\lim _{n\rightarrow \infty} \frac{1}{n}H(X^n),
\end{align}
where $X^n=(X_1,X_2\hdots,X_n)$. Thus 
$\bar{H}(X)$ measures the average asymptotic uncertainty per symbol \cite{cover2012elements}. 
\begin{figure*}[ht!]
\centering
\includegraphics[width=\textwidth]{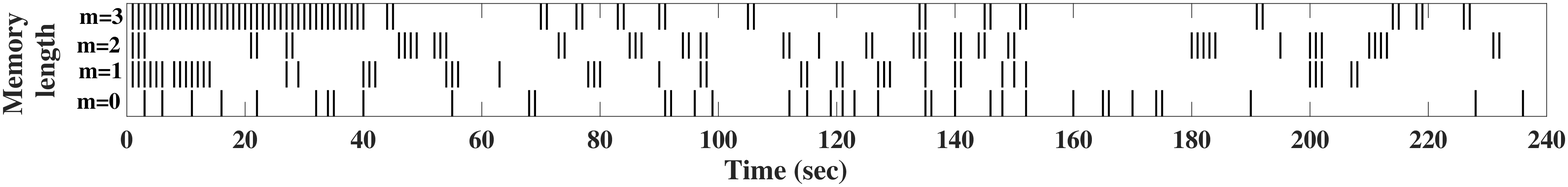}
\vspace{-2em}
\caption{Raster plots of $m$-th ($m=0,1,2,3$) order sources given in the example in Section \ref{ssec:simulation}.}
\label{fig:Rasterplot_ex2} 
\end{figure*}

 \begin{figure*}[t!]
\centering
\begin{tabular}{cccc}
\multicolumn{4}{c}{
\includegraphics[width=15cm]{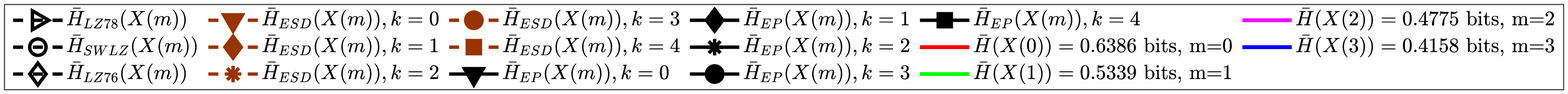}
}
\vspace{-0.9em}
\\ 
\hspace*{-3.2em}

\includegraphics[width=4.8cm,height=0.17\textheight]{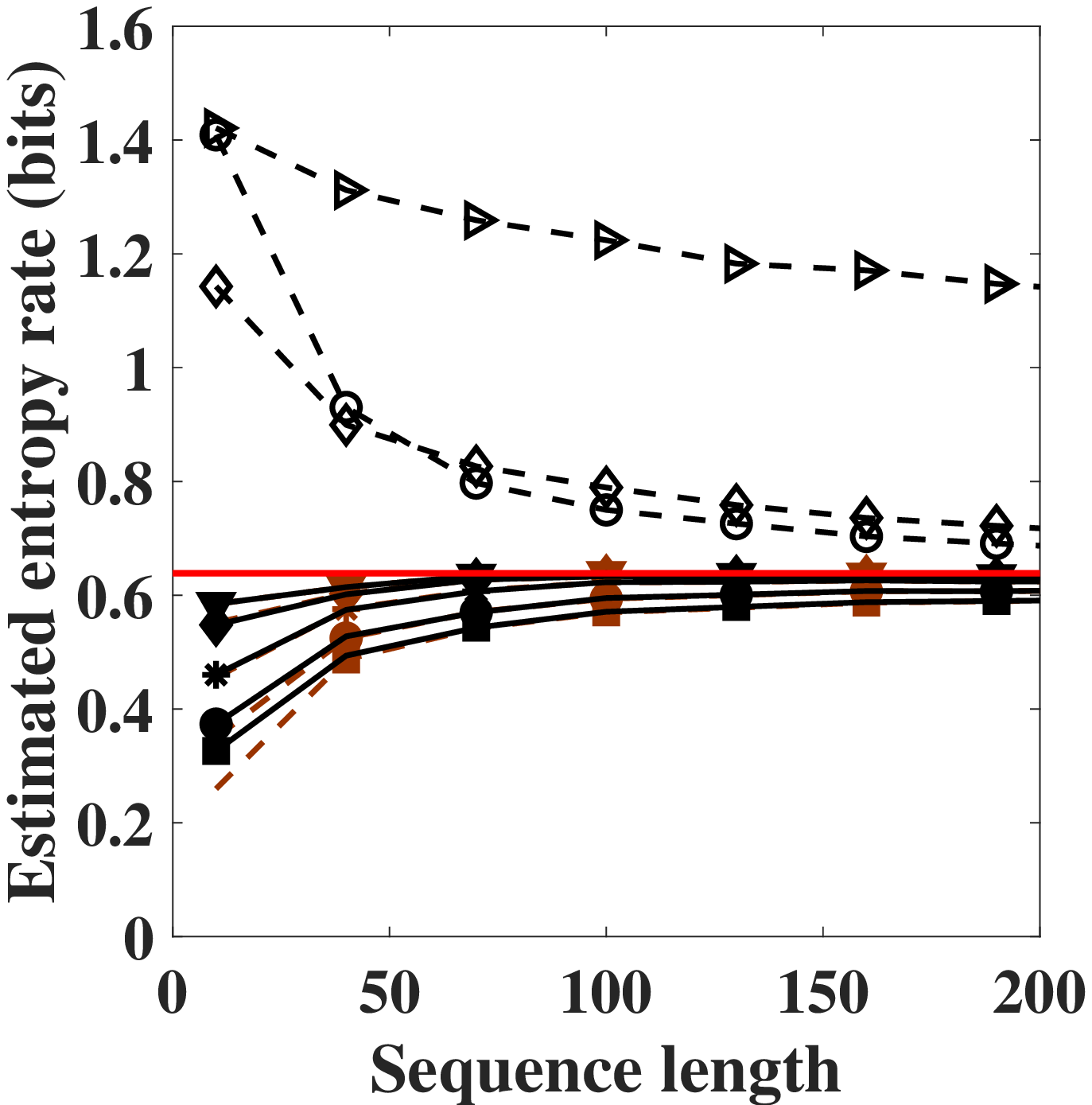}
\hspace*{-4.5em}
&

\includegraphics[width=4.8cm,height=0.17\textheight]{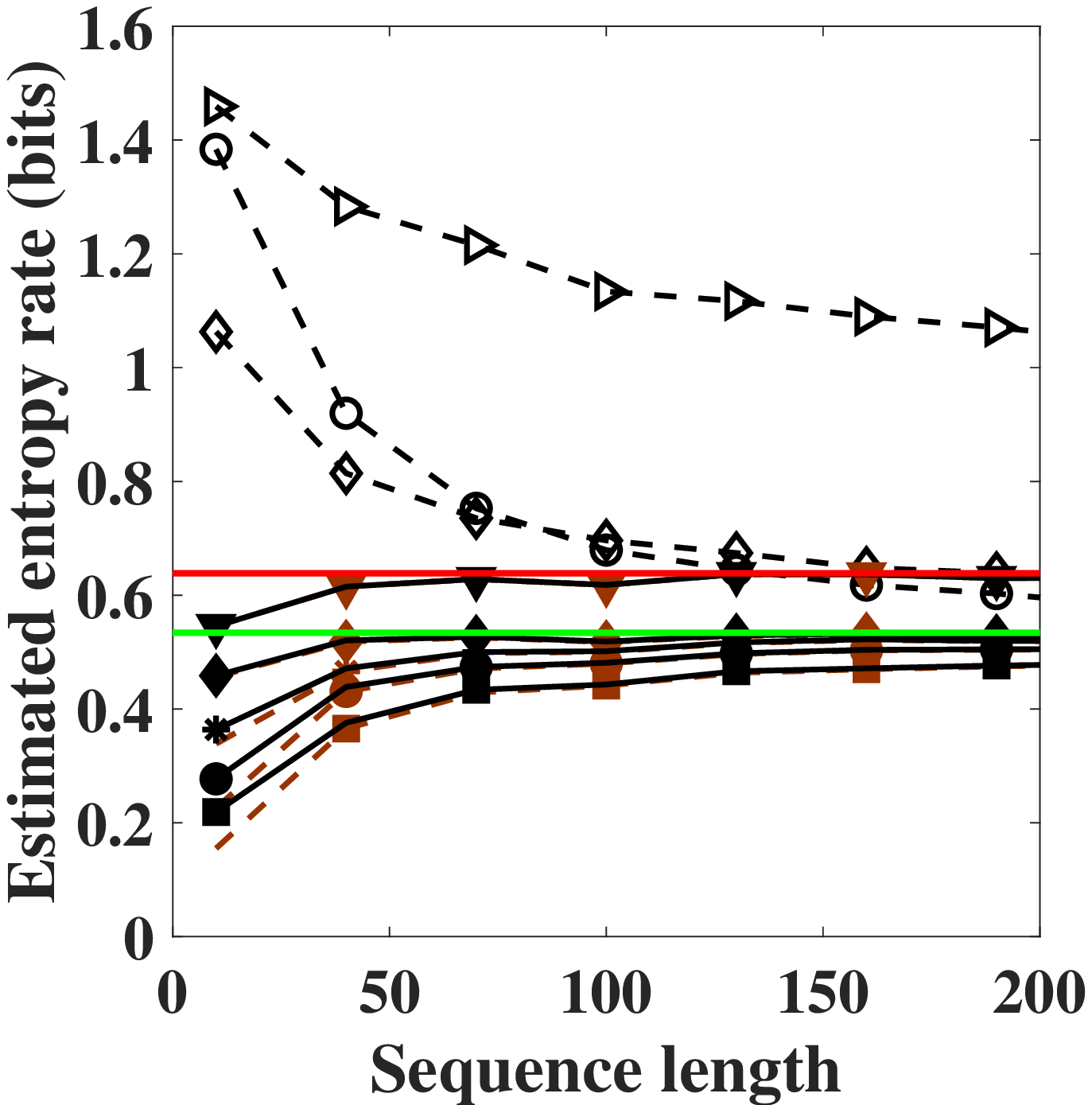}
\hspace*{-2.35em}
&
\includegraphics[width=4.8cm,height=0.17\textheight]{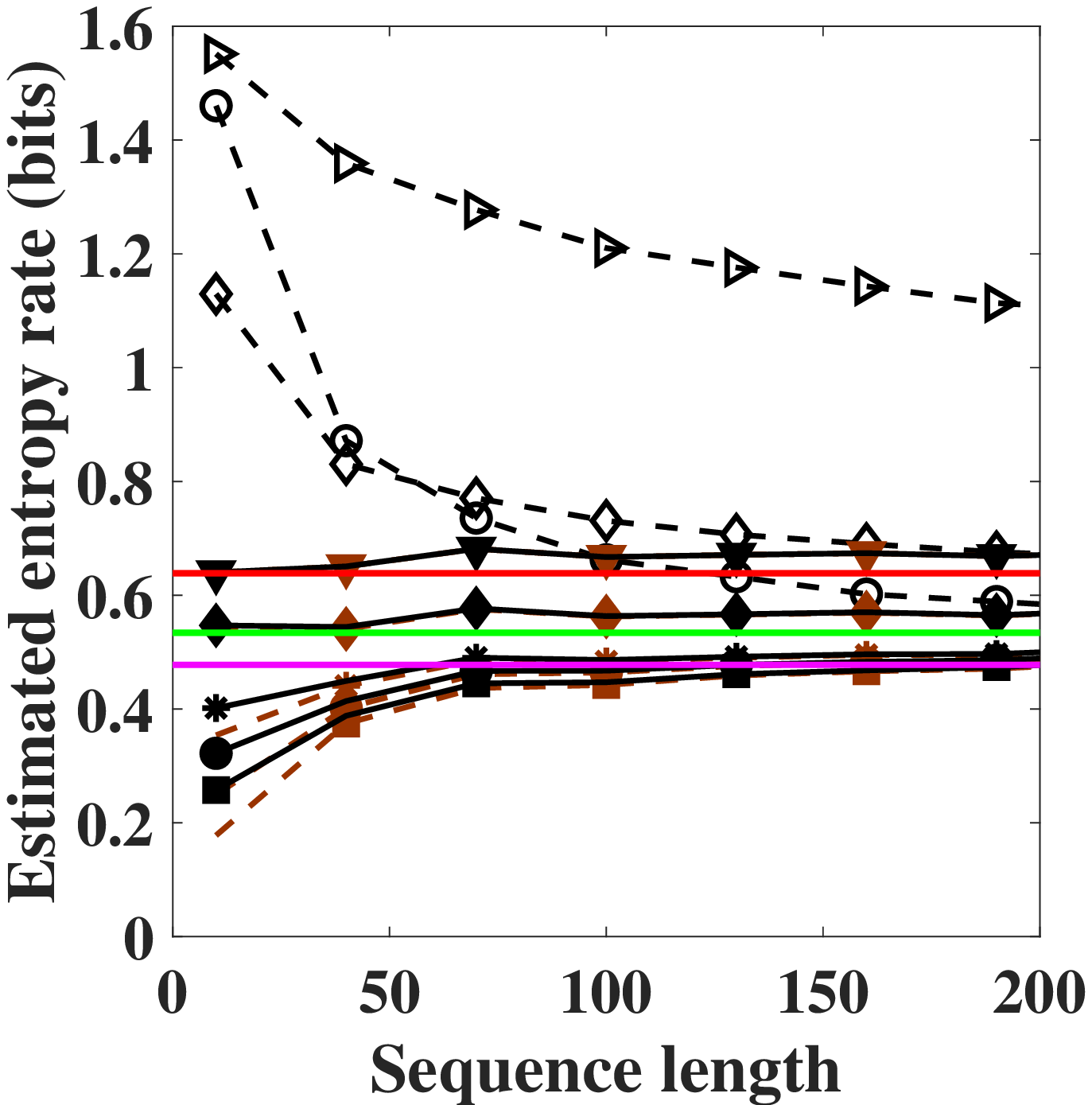}
\hspace*{-2.35em}
&
\includegraphics[width=4.8cm,height=0.17\textheight]{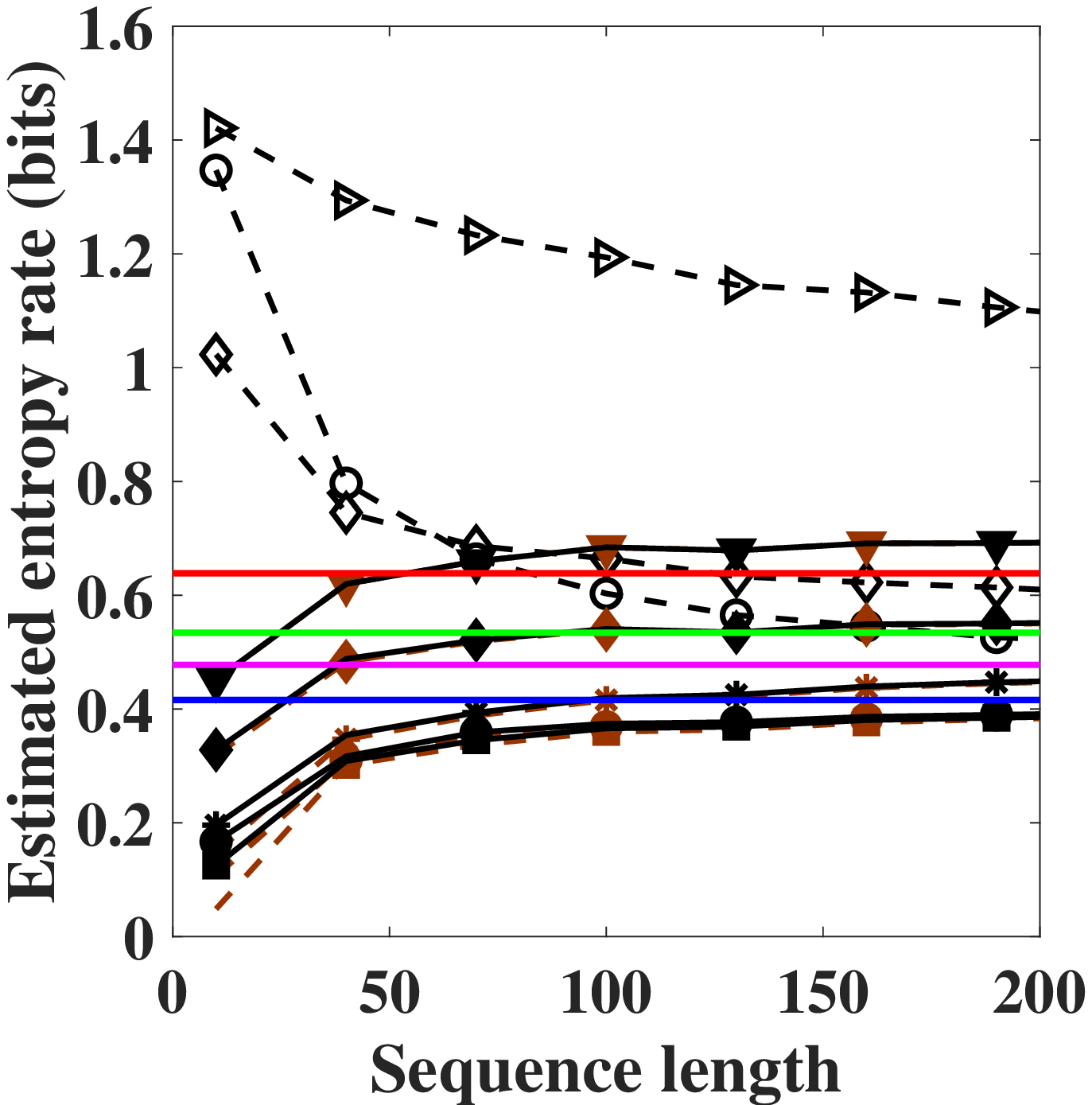}
\vspace*{-0.5em}
\\
\footnotesize{(a) $m=0$,  $\bar{H}(X(0))= 0.6386 $ bits.}
&
\footnotesize{(b) $m=1$, $\bar{H}(X(1))= 0.5339$ bits.}
&
\footnotesize{(c) $m=2$, $\bar{H}(X(2))= 0.4775$ bits.} 
&
\footnotesize{(d) $m=3$, $\bar{H}(X(3))= 0.4158$ bits.}
\end{tabular}
\caption{Comparison of the proposed entropy rate estimate $\bar{H}_{EP}(X)$ with estimates based on LZ variants $\bar{H}_{LZ78}(X)$, $\bar{H}_{SWLZ}(X)$, $\bar{H}_{LZ76}(X)$, and that based on empirical stationary distribution $\bar{H}_{ESD}(X)$ with context lengths $k=0,1,2,3,4$ for $m$-th order Markov sources: (a) $m=0$,  $\bar{H}(X(0))= 0.6386 $ bits; (b) $m=1$, $\bar{H}(X(1))= 0.5339$ bits; (c) $m=2$, $\bar{H}(X(2))= 0.4775$ bits; (d) $m=3$, $\bar{H}(X(3))= 0.4158$ bits.}
\label{fig:simulation2_m1234}
\end{figure*}

\subsubsection{
Lempel-Ziv estimators of entropy rate}
\label{sssec:LZ}
Variants of Lempel-Ziv (LZ) algorithms have been proposed for (text) compression, which also estimate the desired entropy rate. In each case, given a stationary and ergodic sequence $X^n=(X_1,X_2\hdots,X_n)$, LZ complexity approaches the entropy rate $\bar H(X)$ with probability one.\\
\underline{LZ-78 algorithm}:
LZ-78 is dictionary based scheme that adds unforeseen subsequences to the dictionary.
Denoting by $c(X^n)$ the number of sub-sequences of a sequence $X^n$ present in the data, the LZ-78 complexity, the average number of bits per symbol, is defined as \cite{cover2012elements}
\begin{align}
K_{LZ78}(X^{n})= \frac{c(X^{n})\log(c(X^{n}))}{n}.
\label{eq:LZ78}
\end{align}
So, the corresponding entropy rate estimate is $\bar H_{LZ78}(X) =\lim_{n\rightarrow \infty} K_{LZ78}(X^{n})$, where one practically uses large $n$ in place of the limit.\\
\underline{Sliding Window Lempel Ziv (SWLZ) algorithm}:
SWLZ is similar to LZ-78, except that  new subsequence is seen at each position $i$ is not appear in previous $i-1$ symbols by sliding a window of length, $L_i$. 
The SWLZ complexity is given by \cite{vegetabile2019estimating}
\begin{align}
K_{SWLZ}(X^n) = \frac{1}{n} \sum_{i=1}^{n} \frac{L_i}{\log n},
\label{eq:LZ77}
\end{align}
So, the corresponding entropy rate estimate is $\bar H_{SWLZ}(X) =\lim_{n\rightarrow \infty} K_{SWLZ}(X^{n})$.\\
\underline{LZ-76 algorithm}:
LZ-76 provides another variation of the basic dictionary-based scheme. The LZ-76 complexity is given by  \cite{amigo2004estimating}
\begin{align}
K_{LZ76}(X^n) = \frac{c(X^n)}{n} \log n, 
\label{eq:LZ76}
\end{align}
where $c(X^n)$ represents total number of sub-sequences. So, the corresponding entropy rate estimate is $\bar H_{LZ76}(X) =\lim_{n\rightarrow \infty} K_{LZ76}(X^{n})$.\\
\underline{Illustration}: Parsing by the aforementioned algorithms are illustrated on an example string $X^n=`0011001010100111$'. \\
{LZ-78}: 0|01|1|00|10|101|001|11  \\
{SWLZ}: 0|01|1|10|0010|010|101|010| \\
101|0100|10011|00111|0111|111|11|1 \\
{LZ-76}: 0|01|10|010|10100|111.
 
\subsubsection{Empirical stationary distribution estimator}
\label{sssec:sdm}
Here, the spiking process is modeled as Markov process of fixed order $k$. The state transition matrix $T$ is estimated from the inferred and thresholded (binarized) spike sequence. Further, the corresponding stationary state distribution $\hat{\pi}$ satisfies $\hat{\pi} = \hat{\pi} T $. The resulting entropy rate estimate $\bar{H}_{ESD}(X)$ is given by \cite{vegetabile2019estimating}
\begin{align}
\bar{H}_{ESD}(X) =  -\sum_{i}\sum_{j} {\hat{\pi}_i}  {T_{ij}} \log {T_{ij}}.
\label{eq:esd}
\end{align}
Refer to columns 8--11 in Table \ref {table:Empirical_Probability_model} for 
the empirical transition probability distribution for $X^n=`01001011010110001100$'. 

\subsection{Proposed Estimator based on Empirical Probabilities}
\label{sssec:proposed}
 we propose an estimator $\bar H_{EP}(X)$ which also models the neuronal spiking process as Markov process up to a prescribed order $k$ as in Section \ref{sssec:sdm} (Table \ref {table:Empirical_Probability_model}). However, we do not use stationary state distribution, and instead use the estimator 
\begin{equation}
\begin{aligned}
\bar{H}_{EP}(X) &= \frac{H(X_1,X_2 \hdots ,X_n)}{n}\\
 &= \frac{1}{n}(H(X_1 )+...+H({X_{k}}\mathbin{\vert}{X_{k-1},X_{k-2},..,X_1}) \\
&  ...+ H({X_{n}}\mathbin{\vert}{X_{n-1},X_{n-2},..,X_1} ))  \quad \quad \text{by chain rule} \\
&= \frac{1}{n}(H(X_1 )+H({X_2}\mathbin{\vert}{X_1} )+.....\\  & + (n-k) H({X_{k+1}}\mathbin{\vert}{X_{k},X_{k-1},..,X_1})) \quad \text{for order $k$}\\ 
& \approx H({X_{k+1}}\mathbin{\vert}{X_{k},X_{k-1},..,X_1})) \quad \text{if $n>>k$},
\end{aligned}
\label{eq: empiricalEntropy} 
\end{equation}
where ($X_k$, $X_{k-1}$ ,....., $X_2$, $X_1$) provides the context of order $k$ (i.e., $k$ previous symbols) for $X_{k+1}$. In the preceding, $H(U\mathbin{\vert}V)$ denotes the conditional entropy of $U$ given $V$.

\section{RESULTS}
\label{sec:results}
 We begin by considering an example Markov process, and turn to neuronal calcium spike trains subsequently.
 \begin{figure}[t!]
\centering  \includegraphics[width=\columnwidth,height=0.13\textheight]{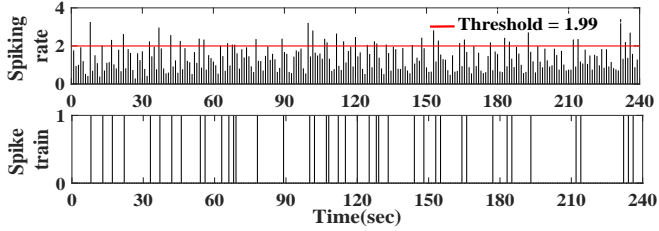} \vspace{-2em}
  \caption{Spike train inferred for neuron 1 from fluorescence intensity data using the aforesaid deconvolution method \cite{vogelstein2010fast} (see top panel in Figure \ref{fig:fluorescence-2}).} 
\label{fig:spiketrain}
\end{figure} 
\textit{\begin{table}[t!]
\centering
\small
\caption{Standard devation of estimated entropy rates for time windows of different lengths for the neuron 4}
\begin{tabular}{ccclclcl}
  \toprule
\multirow{2}{*}{\begin{tabular}[c]{@{}c@{}}\\ \\ Window \\ length \\  \bottomrule \end{tabular}} & \multicolumn{5}{c}{Standard deviation} \\  \bottomrule                                                \\  
 & \begin{tabular}[c]{@{}c@{}}  $\bar H_{SWLZ}(X) $ \\  \\  \bottomrule \end{tabular} & \multicolumn{6}{c}{\begin{tabular}[c]{@{}c@{}}Empirical probability based \\ estimate, $\bar H_{EP}(X) $\\  \bottomrule  \end{tabular}} \\
 &  & \multicolumn{2}{c}{k=0}   & \multicolumn{2}{c}{k=1}   & \multicolumn{2}{c}{k=2}  \\  \bottomrule
40  & 0.0500 & \multicolumn{2}{c}{0.0741}  & \multicolumn{2}{c}{0.0834}  & \multicolumn{2}{c}{0.0712} \\
80  & 0.0190 & \multicolumn{2}{c}{0.0710}      & \multicolumn{2}{c}{0.0745} & \multicolumn{2}{c}{0.0621} \\
120 & 0.0126 & \multicolumn{2}{c}{0.0649}  & \multicolumn{2}{c}{0.0640}      & \multicolumn{2}{c}{0.0530} \\
160  & 0.0124 & \multicolumn{2}{c}{0.0450}      & \multicolumn{2}{c}{0.0495}& \multicolumn{2}{c}{0.0475} \\
200 & 0.0120 & \multicolumn{2}{c}{0.0394}    & \multicolumn{2}{c}{0.0343}      & \multicolumn{2}{c}{0.0422} \\
 \bottomrule
 \end{tabular}
  \label{table:Sd_windows_k012}
\end{table}
}
\begin{figure}[t!]
  \centering
\subfigure{
 \includegraphics[width=0.8\columnwidth]{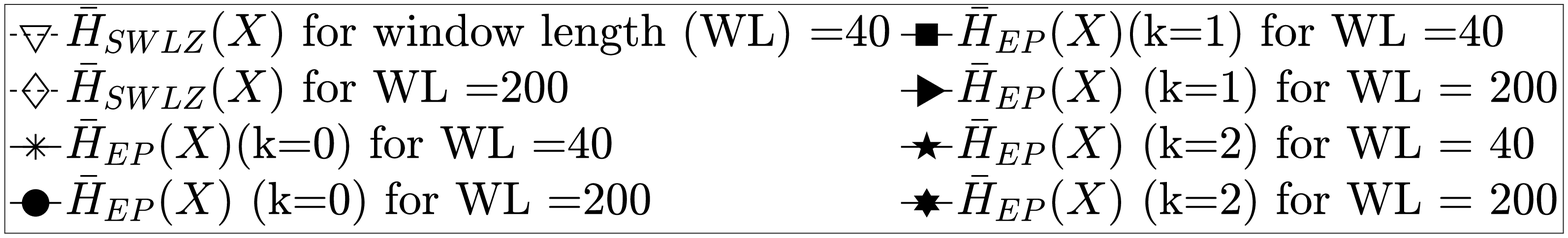}
}\vspace{-1.2em}\\
\centering
\subfigure{
  \includegraphics[width=0.6\columnwidth,,height=0.18\textheight]{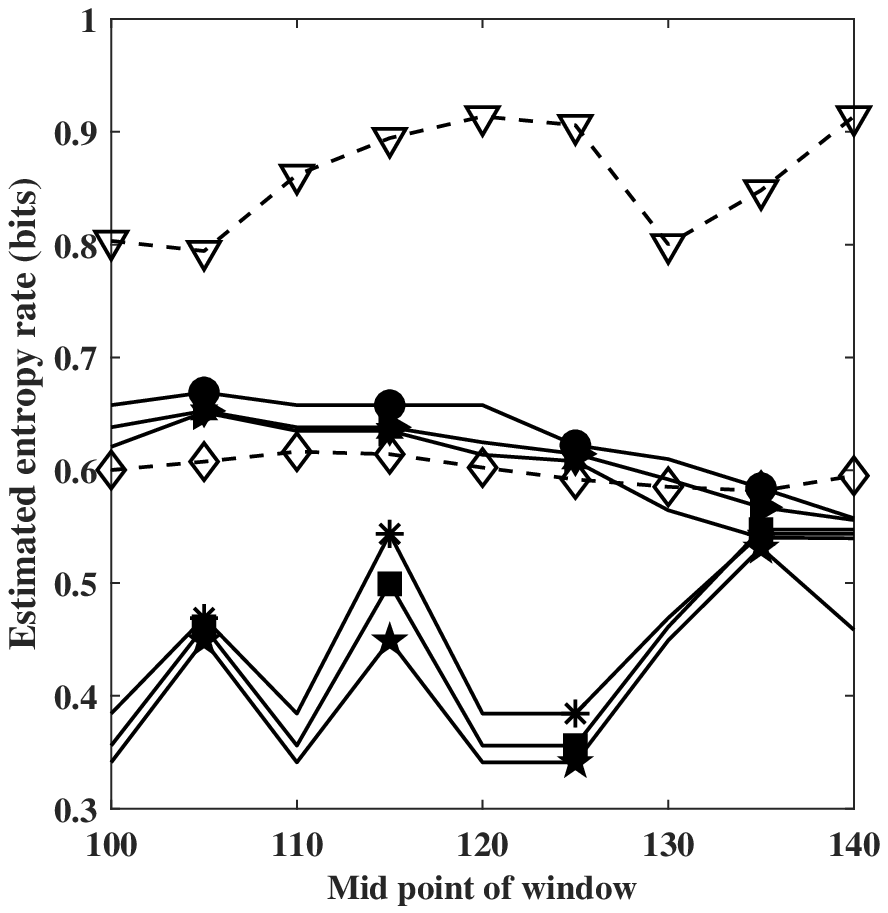}} \vspace{-1em}
 \caption{Variation in the proposed entropy rate estimate $\bar{H}_{EP}(X)$ and that in the competing estimate $\bar{H}_{SWLZ}(X)$ with context lengths $k= 0,1,2$ against different window lengths considering neuron 1.}
\label{fig:estimates_windows_k012}
\end{figure}


\begin{figure*}[t!]
\centering
\includegraphics[width=\textwidth,height=0.07\textheight]{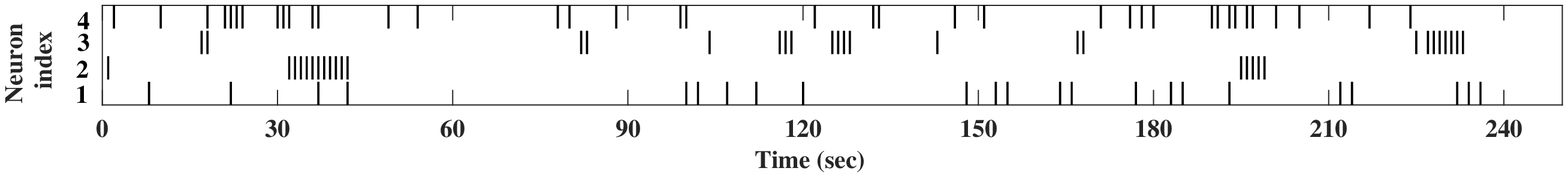}
 \vspace{-2.5em}
\caption{Raster plots of $m$-th ($m=0,1,2,3$) order sources given in the example (Section \ref{ssec:simulation}).}
\label{fig:Rasterplot_ex2}
\end{figure*}

\begin{figure*}[t!]
\centering
\begin{tabular}{cc}
\includegraphics[width=0.9\textwidth]{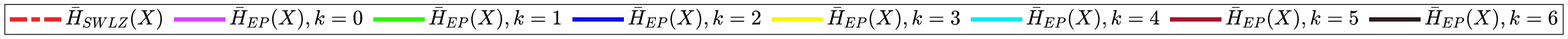}\vspace{-0.9em}\\
\subfigure[]{
\includegraphics[width=4.2cm,height=0.12\textheight]{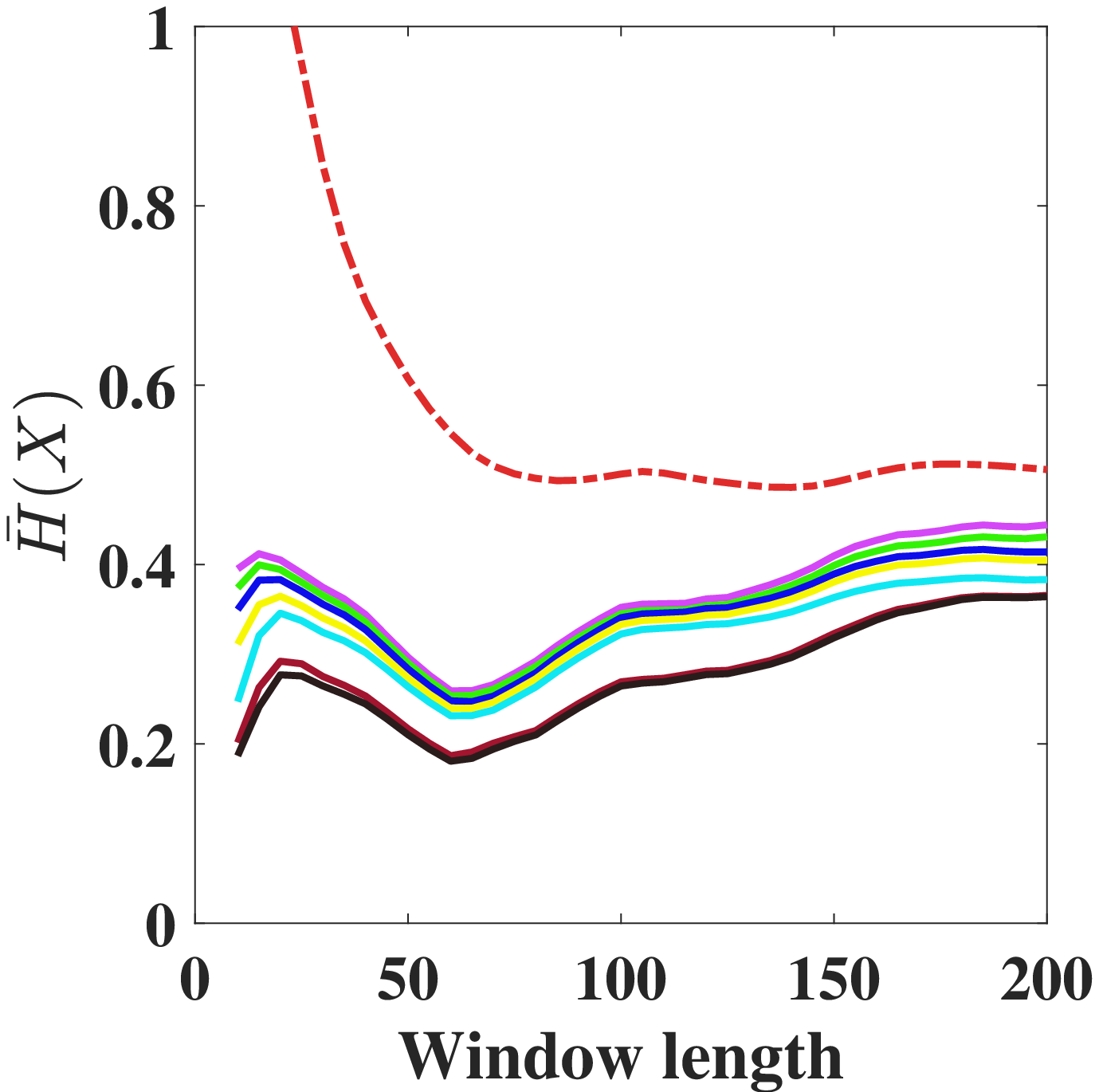}
\label{fig:nd1}\hspace{-1em}}
\subfigure[]{
\includegraphics[width=4.2cm,height=0.12\textheight]{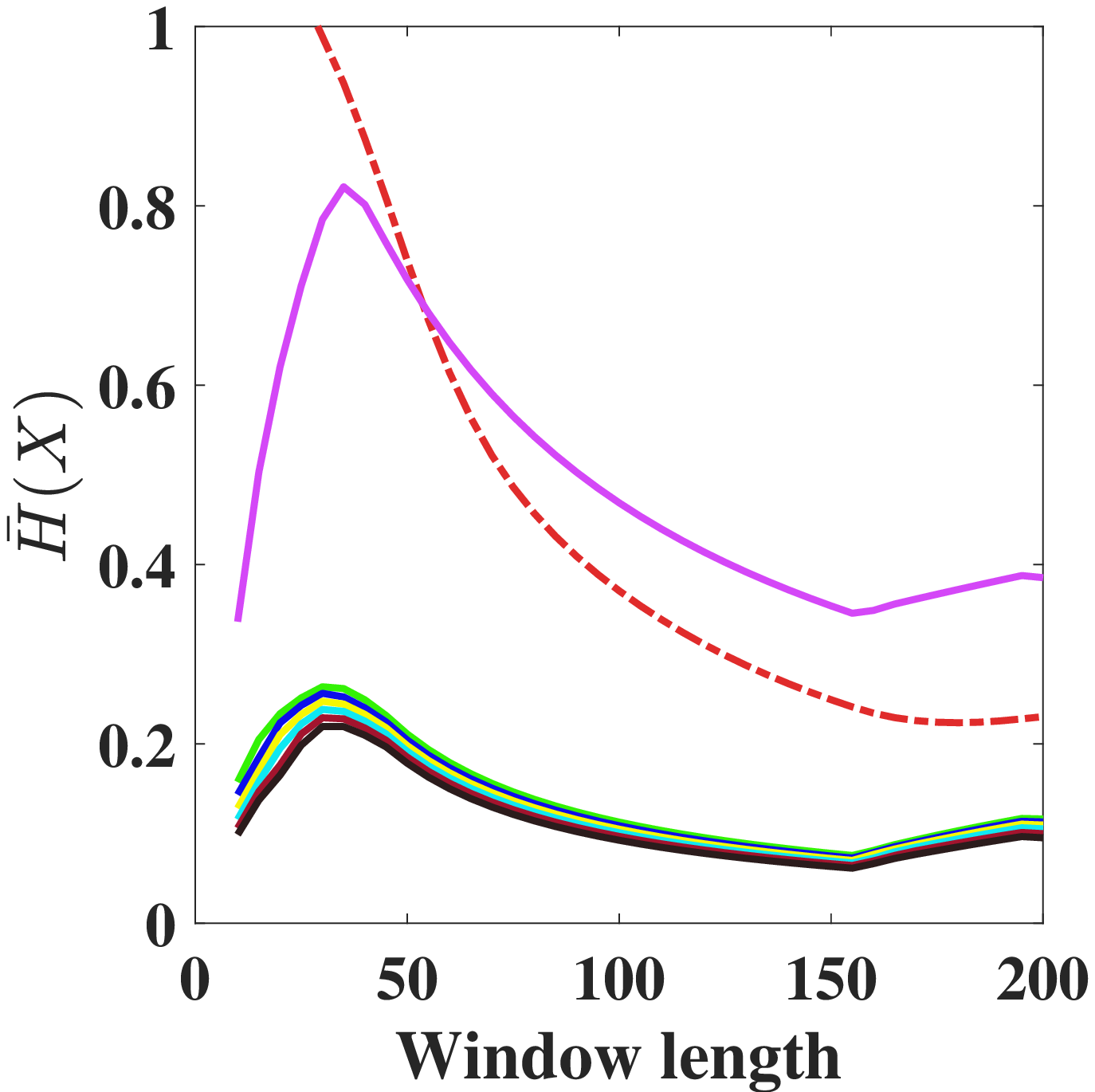}
\label{fig:nd2}\hspace{-1em}} 
\subfigure[]{
\includegraphics[width=4.2cm,height=0.12\textheight]{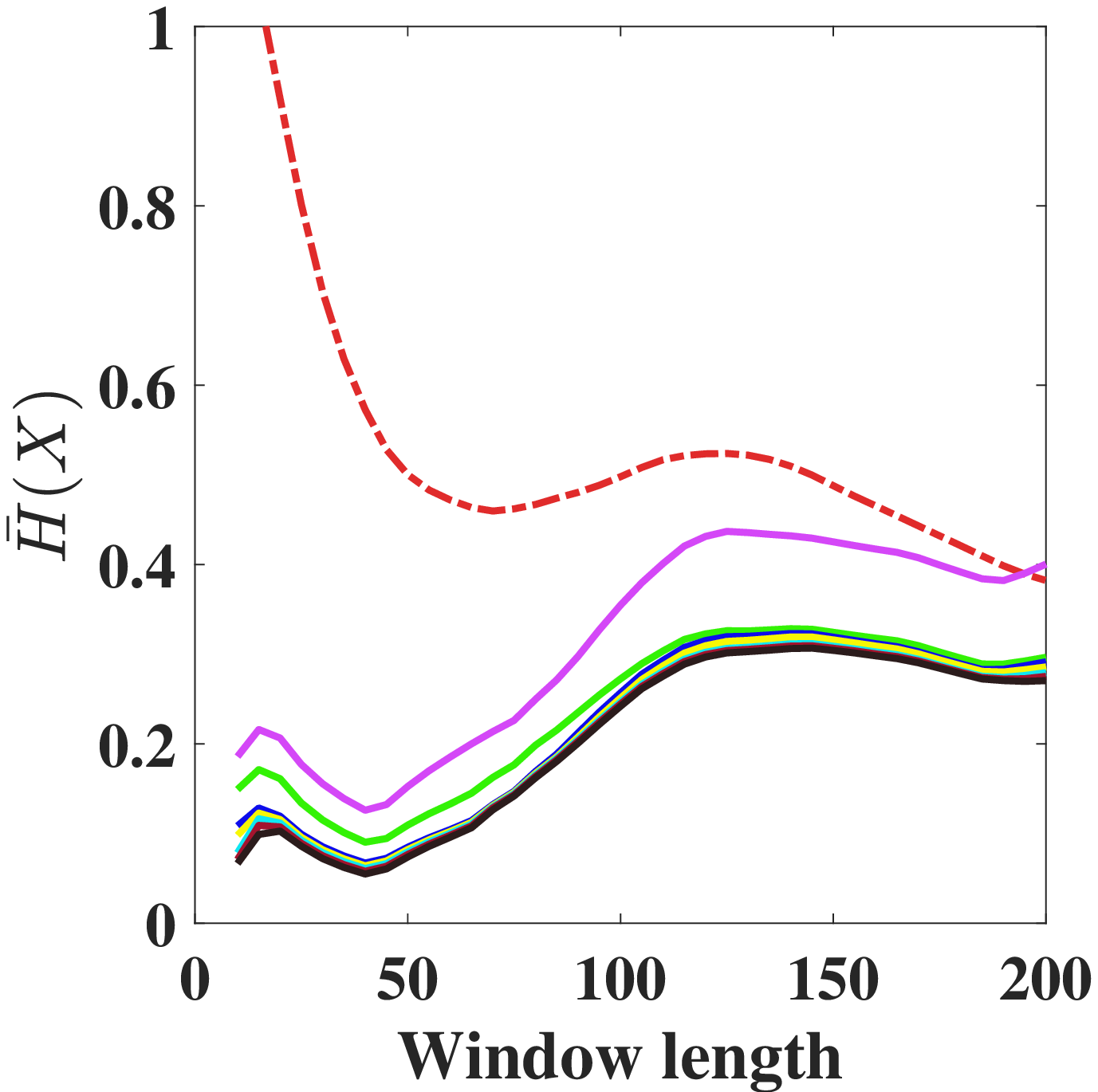}
\label{fig:nd3}\hspace{-1em}} 
\subfigure[]{
\includegraphics[width=4.2cm,height=0.12\textheight]{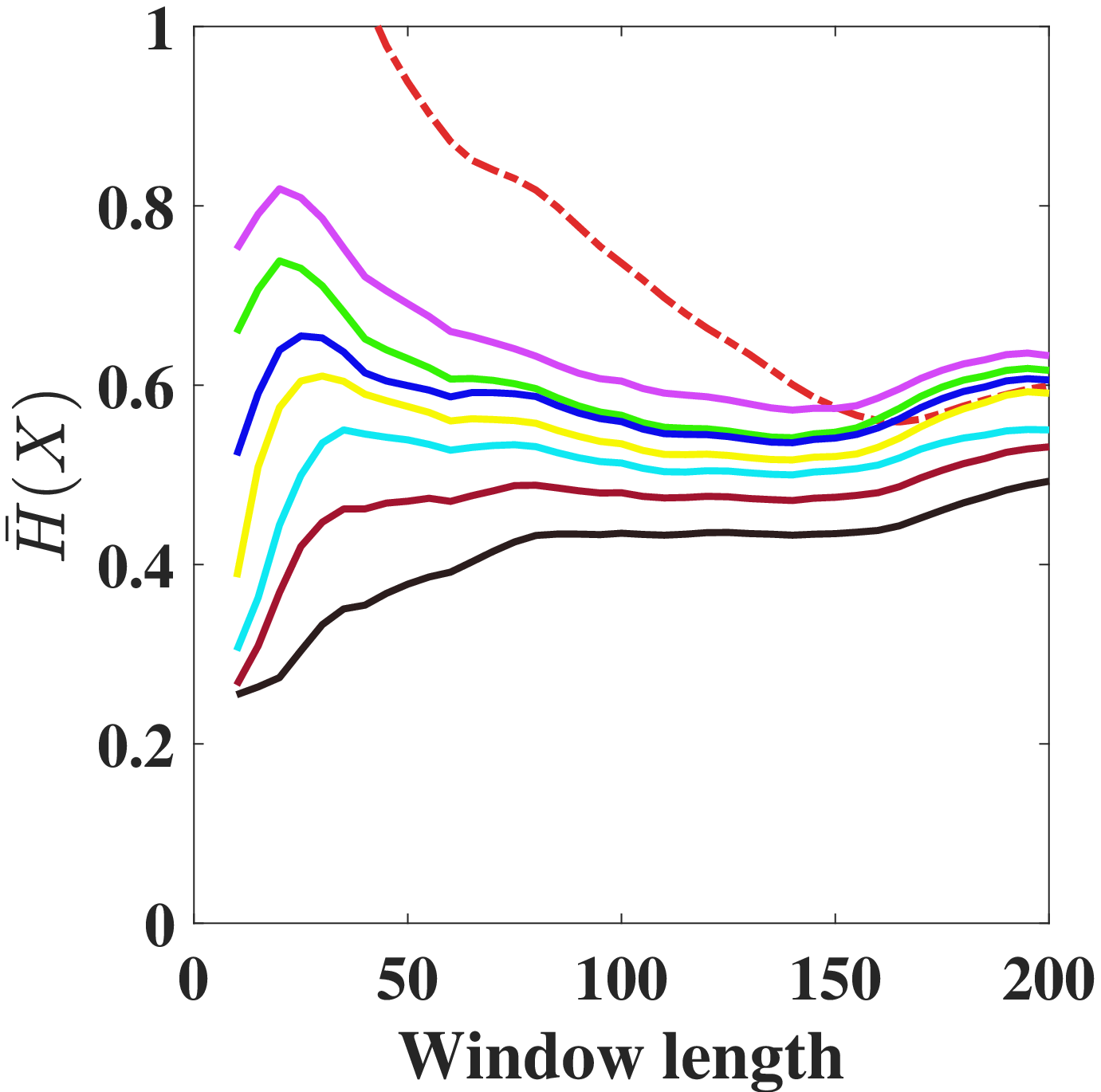}
\label{fig:nd4}}
\end{tabular} \vspace{-0.5em}
\caption{Temporal variation in mean of entropy rate estimates based on SWLZ algorithm and empirical probability based methods for $k=0-6$: (a),(b),(c),(d) for neurons 1, 2, 3, 4, respectively}
\label{fig:TempVariation}
\end{figure*} 
\subsection{Results for Example Markov Process}
\label{ssec:simulation}
One can specify a Markov process of order $m$ by the $m$-th order probability transition matrix $T_m$ and the $m$-th order marginal distribution $\pi_m$, which in turn specify $m$-th order joint distribution \cite{cover2012elements}. various lower order joint distributions can be obtained via marginalization
\begin{align}
 P( X_{n-1}, \hdots ,X_{n-(m-i)}) &=\hspace{-1.3em} \sum_{x_{n-(m-i+1)}}\hspace{-1.3em}P( X_{n-1}, \hdots, X_{n-(m-i+1)}) 
 \label{eq:sp}
\end{align}
while varying $i$, and the corresponding transition probabilities follow from 
\begin{align}
 P(X_n \vert X_{n-1}, \hdots ,X_{n-(m-i)}) &= \frac{P(X_n, X_{n-1}, \hdots ,X_{n-(m-i)})}{ P( X_{n-1}, \hdots ,X_{n-(m-i)})}. 
 \label{eq:tp}
\end{align}
We considered an example process $X(3)$ (the number in parenthesis indicating the order) of order $m=3$ having probability transition matrix
\newcommand{\matindex}[1]{\mbox{\scriptsize#1}}
\[\scriptstyle
  T_3=\begin{blockarray}{ccccccccc}
  \small 
\matindex{000} & \matindex{001}& \matindex{010} & \matindex{011}  & \matindex{100}   &  \matindex{101} & \matindex{110} & \matindex{111} & \\
    \begin{block}{(cccccccc)c}
     0.04& 0.95 & 0.88 &0.92&0.99 & 0.93 & 0.09 &0.9 & \hspace{-0.7em}\matindex{0} \\
  0.96 & 0.05&0.12& 0.08 & 0.01 & 0.07& 0.91& 0.1& \hspace{-0.7em}\matindex{1}\\
    \end{block}
  \end{blockarray}
\]
and marginal joint distribution 
 $$\small \mathbf{\pi_3}=\begin{bmatrix} 0.02 & 0.07  & 0  & 0.07 & 0.06& 0.01&0.07& 0.70 \end{bmatrix}.$$
Entropy rate of this process $\bar H(X(3))=$0.4158 bits. One may compute 
the lower order transition matrices $T_2$  $T_1$ and corresponding marginal joint distributions $\pi_2$ and $\pi_1$ 
using (\ref{eq:sp}) and (\ref{eq:tp}). In the same vein, $\pi_0$ may also be computed. Thus, we obtain processes $X(2)$, $X(1)$ and $X(0)$ which are consistent with the original process $X(3)$. We compute $\bar H(X(2))=$0.4775 bits, $\bar H(X(1))=$0.5339 bits and $\bar H(X(0))=$0.6386 bits.
The raster plot for a realization of each of the above processes is shown in Fig. \ref{fig:Rasterplot_ex2}.

We next demonstrate the efficacy of the proposed estimator. Specifically, we calculated $\bar H_{EP}(X(m))$ using (\ref{eq: empiricalEntropy}), for each of the context lengths $k=0,1,2,3$ and each of the generated processes $X(0),X(1),X(2),X(3)$ (corresponding to $m=0,1,2,3$, respectively), and depict those against the window (sequence) length in Figures \ref{fig:simulation2_m1234}(a)-(d), respectively. In the respective figures, we compared the proposed method 
against various LZ estimates, $\bar H_{LZ78}(X)$, $\bar H_{SWLZ}(X)$, and $\bar H_{LZ76}(X)$, as well as the approximate stationary distribution estimate $\bar H_{ESD}(X)$ (see (\ref{eq:LZ78}), (\ref{eq:LZ77}), (\ref{eq:LZ76}) and (\ref{eq:esd}), respectively). Our method
tends to converge to $\bar H(X(\min(m,k))$, for each pair $(k,m)$, as designed.
In particular, we overestimated entropy rate when the order was underestimated, i.e., $k < m$. For instance, in Figure $5(c)$, while $\bar H(X(2))
=$0.4775 bits for $m=2$, we overestimate $\bar H_{EP}(X(2))\approx$ 0.65 bits for $k=0$ and $\approx$ 0.59 bits for $k=1$. 
However, when $k\ge m$, the proposed estimator was accurate, and the convergence to $\bar H(X(m))$ was faster than the each of the reference algorithms. 
For instance, in Figure $5(a)$, $\bar H_{EP}(X(0))$ ($m=0$)  approximately converges to $\bar H(X(0))=$ 0.6386 bits for each of $k=0,1,2,3$, faster than reference estimates. For sequence/window lengths below 50, $\bar H_{ESD}(X)$ underestimates more than the proposed 
estimator $\bar H_{EP}(X)$, but is otherwise essentially identical, and will not be considered henceforth. Among the rest, the nearest competitor is $\bar H_{SWLZ}(X)$, which shows the fastest convergence among LZ variants, and will be used for comparison in subsequent analysis.

 \subsection{Experimental Results}
\label{ssec:experimental}

For experimental analysis, we chose four neurons, indexed 1, 2, 3, 4 (see Fig.\ref{fig:Calcium Imaging}) whose calcium responses are shown in Fig. \ref{fig:fluorescence-2}. As explained earlier, We first applied the deconvolution method to infer spike trains, and computed the spiking rates and then spiking threshold $P_{th}=1.99$ (using (\ref{eq:threshold})) to remove spurious spikes. In Fig. \ref{fig:spiketrain}, we showed the inferred spike train for neuron 1, and analogous binary spike sequences were used for further analysis.

\subsubsection{Optimization of block/window length}
\label{sssec:stationarity}
We studied the behavior of the proposed estimator $\bar H_{EP}(X)$ for neuron 1 for different lengths of time windows, and compared it with $\bar H_{SWLZ}(X)$, as stated earlier. For that purpose, the midpoint of all time windows of length 40, 80, 120, 160, 200 was fixed at 100 and shifted by 5 every time up to 140. Here, we varied the context length within $k=0,1,2$. Subsequently, we estimated entropy rates in the aforementioned time windows for different $k$, whose standard deviations  are furnished in Table~\ref{table:Sd_windows_k012}. We observed that entropy rate estimates had smallest standard deviation for windows of length 200 for each $k$. Such behaviour was clearly corroborated in Figure \ref{fig:estimates_windows_k012}. Note that $\bar H_{SWLZ}(X)$ exhibited less deviation in windows of length 200; however, it also always overestimated the entropy rate, and proved unsuitable for present comparison. Thus, the window length of 200 is most well suited for entropy rate estimation. However, such analysis should be carried out with more neurons for more comprehensive conclusions.  

  \subsubsection{Temporal variation in entropy rate estimates}
\label{sssec:Temporal}
We next turn to studying temporal variation in information content. To this end, we again considered with aforementioned neurons $1,2,3,4$, and the spike sequences are shown in raster plot in Fig. \ref{fig:Rasterplot_ex2}. Specifically, we  plotted temporal variation of mean estimated entropy rates of aforementioned sequences using  $\bar H_{EP}(X)$ and $\bar H_{SWLZ}(X)$ in Fig. \ref{fig:TempVariation}. Here, the context length $k$ was varied from 0 to 6. The entropy rate estimates, $\bar H_{EP}(X)$ converges to a value between sequence lengths, 150 and 200 unlike $\bar H_{SWLZ}(X)$. For neuron 1, the entropy rate estimates, $\bar H_{EP}(X)$  were almost same across time for the orders $k=0,1,...,5$ unlike for $k=6$, for which slight underestimation is observed (Fig. \ref{fig:nd1}). Further, there was large gap in estimates, $\bar H_{EP}(X)$, $k=0$, and $\bar H_{EP}(X)$, $k=1,...,6$ for neurons 2 and 3 as shown in Figures \ref{fig:nd2} and \ref{fig:nd3}.  
This suggested that there could be process memory of order $k= 1$ for neurons 2 and 3 unlike neuron 1, for which order $k=0$ appears enough. However, the temporal behaviour of neuron 4 was completely different from the aforementioned neurons as shown in Figure \ref{fig:nd4}. The proposed entropy rate estimates $\bar H_{EP}(X)$ were different for each $k$ and those were overestimated for $k=0,1,2$ as compared to $\bar H_{SWLZ}(X)$. This suggested that there could be complex process memory structure for this type of neurons. In other words, we clearly observed heterogeneity in entropy rate estimates, considering only four neurons. We believe that the proposed estimator can provide a useful tool for studying heterogeneity in process memory in large neuron populations.

\section{CONCLUSION}
\label{sec:conclusion}
In this paper, we proposed a empirical probability method for estimating entropy rate to quantify the information content and study its temporal variation in calcium spike trains of hippocampal neurons. We demonstrated our method with synthetic Markov sequences as well as experimental spike sequences and found to have fast convergence in short time windows compared to other methods. Although, this method is able to identify the window for slow temporal statistics and capture the heterogeneity in information estimates, there remains much room for improvement. Importantly, our analysis has to be thoroughly investigated for large number of neurons and neuron populations. Further, complex process memory structure of neurons requires deeper analysis. The present entropy rate analysis could be useful for understanding neuronal information encoding under normal and diseased conditions, and identifying disease signatures. Our tool also enables potential investigations into correlation between neuronal information content and other physiological aspects, such as neuronal activity \cite{swain2018confocal}, as well as synchronicity \cite{swain2018sync}.

\balance
  
\section{ACKNOWLEDGMENT}
We thank Drs. Mennerick and Gautam for providing materials and equipment. Sathish Ande thanks the Ministry of Electronics and Information Technology (MeitY), the Government of India, for fellowship grant under Visvesvaraya PhD Scheme. 

\bibliographystyle{IEEEtran}
\bibliography{ReferencesBib}

\end{document}